\documentstyle[preprint,aps]{revtex}
\begin{document}
\title{Phase separation and vortex states in binary mixture of
Bose-Einstein condensates}
\author{ S. T. Chui$^a$, V. N. Ryzhov$^b$ and E. E. Tareyeva$^b$}
\address{a: Bartol Research Institute, University of Delaware, Newark, DE
19716\\b:Institute for High Pressure Physics, Russian Academy of
Sciences, 142 190 Troitsk, Moscow region, Russia}
\maketitle

\begin{abstract} The phase separation and vortex states
in two-component Bose-Einstein condensate consisting of
$|F=1,m_f=-1>$ and $|2,1>$ internal spin states of $^{87}Rb$
atoms are considered in the framework of Thomas-Fermi approximation.
It is shown that in nonrotating system the atoms in the state
$|1,-1>$ form a shell about the atoms in the state $|2,1>$.
The critical angular velocity for each state is calculated.
These velocities  depend drastically on the relative
concentrations of the components, the critical angular
velocity of the outer component being less than the angular
velocity of the inner one. It is shown that the atoms in the
$|1,-1>$ state can form a rotating ring about the resting core
of the atoms in the state $|2,1>$.
\end{abstract}

\bigskip


\bigskip

        The realization of Bose-Einstein Condensation (BEC)
in dilute atomic gases offers new
opportunities for studying quantum degenerate fluids \cite{[1]}.
These condensates, which contain thousands
of atoms confined to microscale clouds, have
similarities to superfluidity and laser, and provide new
testing ground for many body physics.

        The modern theoretical description of dilute BEC
originates from Bogoliubov's seminal
1947 paper where he showed that weak repulsive
interaction qualitatively change the excitation spectra
from quadratic free particle form to
a linear phonon-like structure. To describe the trapped condensates
at $T=0$ one can use the Gross-Pitaevskii (GP)
(nonlinear Schrodinger) equation for the condensate wave
function \cite{[2]}. This equation appears as
the generalization of the Bogoliubov theory for the
inhomogeneous phase. It was widely used to discuss
the ground state properties and collective excitations in BEC.

        Bulk superfluids are distinguished from normal
fluids by their ability to support dissipationless
flow. This ability is closely related to the
existence of stable quantized vortices. Such vortices have been
widely studied in superfluid $^4He$. Recently
clear
experimental evidences of the existence of a
vortex in trapped BEC \cite{bose_vor,wilhol,madison} was reported.
Unlike superfluid helium,
the trapping potential makes alkali BEC
nonuniform. Theoretical work has been concentrated  on
the critical angular velocity of the vortex creation,
the collective excitations of BEC in the presence
of the vortex and considerations of stability of the vortex
\cite{[3]}-\cite{[15]}.
It was shown that unlike superfluid helium
where the vortex is locally stable, for weakly
interacting gases stable quantized vortices exist
only in a driven system, and become unstable without
imposed rotation. So this system
cannot be considered a superfluid \cite{[3],[7],[8]}.

        In the study of quantum fluids, the most
interesting behavior has been found in the behavior
of fluid mixtures. At the present time,
two experimental groups have observed trapped multiple
condensates as realized in a magnetic trap in rubidium \cite{[10]}
and in an optical trap in sodium \cite{[11]}. In
these experiments, the spatial separation of condensates
has been observed. One can distinguish two
types of spatial separation: (a) potential separation,
caused by external trapping potentials; (b) phase
separation, which can occur in the absence of external
potentials due to the interaction between two
condensates. It is the latter type of phase separation that
we consider in this paper. This type of phase separation
has been observed in the experiments on the
simultaneously trapped condensates, consisting of the $^{87}Rb$
atoms in the $|2,1>$ and $|1,-1>$  spin states
(states 2 and 1 correspondingly) \cite{[10]}.
In this case the intraspecies and interspecies scattering lengths
denoted correspondingly as $a_{11}, a_{22}, a_{12}$
are in the proportion $a_{11} :a_{12} :a_{22} =1.03:1:0.97$
with the
average of the three being $55(3)\AA$ \cite{[10],[12]}.

In this paper we consider the behavior of binary mixture of
Bose-Einstein condensates of alkali atoms.
We calculate the critical angular velocity
needed to create stable vorticies in either component
in the rotating frame.
This quantity is of crucial importance in view of
the experimental possibility of creating vortices
by rotation of the confining trap \cite{madison}.

        The physics of interpenetrating  Bose fluids is very
rich and far from a complete understanding.
While properties of  a rotating single component
Bose fluids are much studied \cite{[3]}-\cite{[15]},
the rotating mixture of  alkali atoms presents a new frontier,
that is essentially virgin territory.

        In order to derive analytic results,
some approximations must be used. A commonly used one
is the Thomas-Fermi Approximation (TFA) ,
which ignores the kinetic energy terms. It has been shown
that in the case of one component condensates the TFA results
agree well  with the numerical
calculations for large particle numbers,
except for a small region near the boundary of the condensate
\cite{[4],[5]}. In fact, even for a small number of particles the
TFA still usually gives qualitatively correct results.
The TFA provides an excellent
starting point of study. However, the TFA should
not be relied upon when a quantitative comparison
of experiment and theory is important. In this case a
numerical approach based on the Monte-Carlo
simulation becomes necessary.

Let us first consider the phase separation in binary mixture
without rotation.

In the case of a two-species condensate, letting $\psi_i({\bf r})$ $(i=1,2)$
be the wave function of species $i$ with particle number $N_i$, we can
write the two coupled nonlinear Schrodinger (Gross-Pitaevskii) equations as:
\begin{eqnarray}
&-&\frac{\hbar^2}{2m_1}\nabla^2\psi_1({\bf r})+
\frac{1}{2}m_1\omega_1^2(x^2+y^2
+\lambda^2z^2)\psi_1({\bf r})-\mu_1\psi_1({\bf r})+ \nonumber \\
&+&G_{11}|\psi_1({\bf r})|^2\psi_1({\bf r})+G_{12}|\psi_2({\bf r})|^2
\psi_1({\bf r})=0; \label{1}\\
&-&\frac{\hbar^2}{2m_2}\nabla^2\psi_2({\bf r})+\frac{1}{2}m_2\omega_2^2(x^2+y^2
+\lambda^2z^2)\psi_2({\bf r})-\mu_2\psi_2({\bf r})+ \nonumber \\
&+&G_{22}|\psi_2({\bf r})|^2\psi_2({\bf r})+G_{12}|\psi_1({\bf r})|^2
\psi_2({\bf r})=0. \label{2}
\end{eqnarray}

Equations (\ref{1}) and (\ref{2}) were obtained by minimization of the
energy functional of the trapped bosons of masses $m_1$ and $m_2$ given by:
\begin{eqnarray}
E(\psi_1,\psi_2)&=&\int\,d^3 r\left[
\frac{\hbar^2}{2m_1}|\nabla\psi_1({\bf r})|^2+\frac{1}{2}m_1\omega_1^2(x^2+y^2
+\lambda^2z^2)|\psi_1({\bf r})|^2+\right. \nonumber\\
&+&\frac{\hbar^2}{2m_2}|\nabla\psi_2({\bf r})|^2+\frac{1}{2}m_2\omega_2^2(x^2+y^2
+\lambda^2z^2)|\psi_2({\bf r})|^2+ \nonumber\\
&+&\left.\frac{G_{11}}{2}|\psi_1({\bf r})|^4+\frac{G_{22}}{2}|\psi_2({\bf r})|^4
+G_{12}|\psi_1({\bf r})|^2|\psi_2({\bf r})|^2 \right]. \label{3}
\end{eqnarray}

The chemical potentials $\mu_1$ and $\mu_2$ are determined by the relations
$\int\,d^3 r|\psi_i|^2=N_i$. The trap potential is approximated
by an effective three-dimensional harmonic-oscillator potential
well, which is cylindrically symmetric about $z$ axis, $\lambda$
being the ratio of angular frequencies in the axial direction
$\omega_{zi}$ to that in the transverse direction
$\lambda=\omega_{zi}/\omega_i$. The experimental value of $\lambda$ is
$\lambda=\sqrt{8}$. The interaction strengths, $G_{11}, G_{22}, G_{12}$ are
determined by the $s$-wave scattering lengths for binary collisions of
like and unlike bosons: $G_{ii}=4\pi\hbar^2a_{ii}/m_i; G_{12}=2\pi\hbar^2a_{12}
/m$, where $m^{-1}=m_1^{-1}+m_2^{-1}$.

Let us consider now the phase separation due to interaction between the two
condensates. In this case
\begin{equation}
\frac{1}{2}m_1\omega_1^2=\frac{1}{2}m_2\omega_2^2. \label{4}
\end{equation}

Let us simplify the equations by using dimensionless variables.
We define the length scale  \begin{equation}
a_{\perp}=\left(\frac{\hbar}{m_1\omega_1}\right)^{1/2}, \label{5}
\end{equation}
and the dimensionless variables
\begin{eqnarray}
{\bf r}&=&a_{\perp}{\bf r}', \label{6}\\
E&=&\hbar\omega_1 E', \label{7}\\
\psi_i({\bf r})&=&\sqrt{N_i/a_{\perp}^3}\psi_i'({\bf r}'). \label{8}
\end{eqnarray}
The wave function $\psi_i'({\bf r}')$ is normalized to $1$.
In terms
of these variables the Gross-Pitaevskii energy functional takes the form:
\begin{eqnarray}
E'&=&\frac{1}{2}\int\,d^3r'\left[N_1|\nabla'\psi_1'|^2+
N_1(x'^2+y'^2+\lambda^2z'^2)|\psi_1'|^2+
N_2\beta^2|\nabla'\psi_2'|^2+N_2(x'^2+y'^2+\lambda^2z'^2)|\psi_2'|^2+ \right.
\nonumber\\
&+&\left.\frac{1}{2}N_1u_1|\psi_1'|^4+\frac{1}{2}N_2u_2\beta^2|\psi_2'|^4
+\frac{2\pi a_{12}}{a_{\perp}}\frac{m_1}{m}
N_1N_2|\psi_1'|^2|\psi_2'|^2\right].
\label{9}
\end{eqnarray}
Here $\beta^2=m_1/m_2=\omega_2^2/\omega_1^2$ and
$u_i=8\pi a_{ii}N_i/a_{\perp}$. In deriving Eq. (\ref{9}) we used
Eq. (\ref{4}). Eqs. (\ref{1}) and (\ref{2}) can be rewritten as:
\begin{eqnarray}
&-&\nabla'^2\psi_1'+(x'^2+y'^2+\lambda^2z'^2)\psi_1'-\mu_1'\psi_1'+
u_1|\psi_1'|^2\psi_1'+ \nonumber\\
&+&\frac{4\pi a_{12}N_2}{a_{\perp}}\frac{m_1}{m}|\psi_2'|^2\psi_1'=0;
\label{10}\\
&-&\beta^2\nabla'^2\psi_2'+(x'^2+y'^2+\lambda^2z'^2)\psi_2'-\mu_2'\psi_2'+
u_2\beta^2|\psi_2'|^2\psi_2'+ \nonumber\\
&+&\frac{4\pi a_{12}N_1}{a_{\perp}}\frac{m_1}{m}|\psi_1'|^2\psi_2'=0;
\label{11}
\end{eqnarray}
where $\mu_i'=2\mu_i/\hbar\omega_1$.

In the TFA, Eqs. (\ref{9}), (\ref{10})
and (\ref{11}) can be further simplified by omitting the kinetic energy.
The phase segregated condensates do not overlap, so we can neglect
the last terms in Eqs. (\ref{9}), (\ref{10}) and (\ref{11}), obtaining from
(\ref{10}) and (\ref{11}), in separate regions that they do not
overlap, simple algebraic equations:
\begin{eqnarray}
|\psi_1'({\bf r}')|^2&=&\frac{1}{u_1}\left(\mu_1'-(\rho'^2+\lambda^2z'^2)\right)
; \label{12} \\
|\psi_2'({\bf r}')|^2&=&\frac{1}{u_2\beta^2}
\left(\mu_2'-(\rho'^2+\lambda^2z'^2)\right)
. \label{13}
\end{eqnarray}
Here $\rho'^2=x'^2+y'^2$. From
Eqs. (\ref{12}) and (\ref{13}) one can see that the condensate density has
the ellipsoidal form.

In the case of phase separation, the energy of the system can be written
in the form
\begin{equation}
E=E_1+E_2, \label{14}
\end{equation}
where
\begin{eqnarray}
E_1&=&\frac{1}{2}\hbar\omega_1 N_1\left[\mu_1'-\frac{1}{2}u_1\int\,d^3r'
|\psi_1'|^4\right], \label{15}\\
E_2&=&\frac{1}{2}\hbar\omega_1 N_2\left[\mu_2'-\frac{1}{2}u_2\beta^2
\int\,d^3r' |\psi_2'|^4\right]. \label{16}
\end{eqnarray}
In order to obtain Eqs. (\ref{15})-(\ref{16}),
Eqs. (\ref{12})-(\ref{13}) have been used.

To investigate the phase separation in the mixture we first
suppose that the condensate 1 atoms form an ellipsoidal shell about
the condensate 2 atoms (we will call this configuration
as the configuration "a"). To determine the position of the boundary between
the condensates, we use the condition of thermodynamic equilibrium
\cite{landau}: the pressures exerted by both condensates must be equal:
\begin{equation}
P_1=P_2. \label{17}
\end{equation}
The pressure is given by \cite{pitaevskii}:
\begin{equation}
P_i=\frac{G_{ii}}{2}|\psi_i|^4. \label{18}
\end{equation}

The condensate 2 has the form of the ellipsoid with long
semiaxis $q$:  \begin{equation} \rho'^2+\lambda^2z'^2=q^2.
\label{19} \end{equation}

From Eqs. (\ref{12})-(\ref{13}) and (\ref{17})-(\ref{19}) one has
the equation for $q$:
\begin{equation}
\mu_1'-q^2=\kappa \mu_2'-\kappa q^2, \label{20}
\end{equation}
where $\kappa=\sqrt{(a_{11}m_2)/(a_{22}m_1)}$.

Chemical potentials $\mu_1'$ and $\mu_2'$ can be obtained using the
normalization conditions
$\int\,d^3r'|\psi_1'|^2=\int\,d^3r'|\psi_2'|^2=1$ and are given
by:
\begin{equation}
\mu_1'=\frac{\mu_1^0}{\left(1-\frac{5}{2}q'^3+\frac{3}{2}q'^5\right)^{2/5}},
\label{21}
\end{equation}
\begin{equation}
\mu_2'=\frac{3}{(\mu_1')^{3/2}q'^3}\left(\frac{2\beta^2(\mu_2^0)^{5/2}}{15}+
\frac{(\mu_1')^{5/2}q'^5}{5}\right), \label{22}
\end{equation}
where  $q=\sqrt{\mu_1'}q'$ and
\begin{equation}
\mu_i^0=\left(\frac{15\lambda u_i}{8\pi}\right)^{2/5}. \label{23}
\end{equation}

From equations (\ref{21})-(\ref{23}) one can determine the chemical
potentials $\mu_1'$ and $\mu_2'$ and the semiaxis of the
phase boundary ellipsoid $q$ as functions of $N_1$ and $N_2$.
The energy of the configuration "a" $E_a=E_{a1}+E_{a2}$ is given by:
\begin{eqnarray}
E_{a1}&=&\frac{1}{2}\hbar \omega_1 N_1 \left\{\mu_1'-\frac{15}{4}
\frac{(\mu_1')^{7/2}}{(\mu_1^0)^{5/2}}\left[\frac{8}{105}-
\left(\frac{q'^3}{3}-\frac{2}{5}q'^5+\frac{q'^7}{7}\right)\right]\right\},
\label{24} \\
E_{a2}&=&\frac{1}{2}\hbar \omega_1 N_2 \left\{\mu_2'-\frac{15}{4}
\frac{(\mu_1')^{3/2}}{\beta^2(\mu_2^0)^{5/2}}\left(\mu_2'^2\frac{q'^3}{3}
-2\mu_2'\mu_1'\frac{q'^5}{5}+\mu_1'^2\frac{q'^7}{7}\right)\right\}.
\label{25}
\end{eqnarray}

Let us now consider the opposite case when
the condensate 2 atoms form an ellipsoidal shell about the condensate 1
atoms (configuration "b"). In this case Eqs. (\ref{20})-(\ref{25})
can be rewritten in the form:
\begin{equation}
\mu_1''-q_1^2=\kappa(\mu_2''-q_1^2),\label{26}
\end{equation}
\begin{equation}
(\mu_2'')^{5/2}=\frac{\beta^2(\mu_2^0)^{5/2}}{1-\frac{5}{2}q_1'^3+
\frac{3}{2}q_1'^5},\label{27}
\end{equation}
\begin{equation}
\frac{15}{2}\frac{(\mu_2'')^{3/2}}{(\mu_1^0)^{5/2}}\left(
\frac{\mu_1''q_1'^3}{3}-\frac{\mu_2''q_1'^5}{5}\right)=1, \label{28}
\end{equation}
\begin{eqnarray}
E_b&=&E_{b1}+E_{b2}, \nonumber\\
E_{b1}&=&\frac{1}{2}\hbar\omega_1 N_1 \left\{\mu_1''-\frac{15}{4}
\frac{(\mu_2'')^{3/2}}{(\mu_1^0)^{5/2}}\left(\mu_1''^2\frac{q_1'^3}{3}-
2\mu_1''\mu_2''\frac{q_1'^5}{5}+\mu_2''^2\frac{q_1'^7}{7}\right)\right\},
\label{29}\\
E_{b2}&=&\frac{1}{2}\hbar\omega_1 N_2\left\{\mu_2''-\frac{15}{4}
\frac{(\mu_2'')^{7/2}}{\beta^2(\mu_2^0)^{5/2}}\left[\frac{8}{105}-\left(
\frac{q_1'^3}{3}-\frac{2}{5}q_1'^5+\frac{q_1'^7}{7}\right)\right]\right\}.
\label{30}
\end{eqnarray}
Here $\mu_1''$ and $\mu_2''$ are the chemical potentials in the
configuration "b", $q_1=\sqrt{\mu_2''}q_1'$ is the long semiaxis
of the boundary ellipsoid, $E_b$ is the energy of the configuration "b".

To estimate which configuration is stable, one has to compare $E_a$ and
$E_b$. Let us first consider the limiting cases  $n_2=N_2/N_1 \ll 1$, and
$n_1=N_1/N_2 \ll 1$.

In the former case $n_2 \ll 1$ the approximate solution of Eqs. (\ref{20})-
(\ref{22}) has the form:
\begin{eqnarray}
q'&=&q_0\left(1+\frac{1}{3}\left(1-\frac{2}{5}\kappa\right)q_0^2-\frac{5}{6}
q_0^3\right),\label{31}\\
\mu_1'&=&\mu_1^0(1+q_0^3), \label{32}\\
\mu_2'&=&\frac{\mu_1^0}{\kappa}\left(1+(\kappa-1)q_0^2+q_0^3\right).
\label{33}
\end{eqnarray}
where
\begin{equation}
q_0=\left(\frac{2n_2}{5\kappa}\right)^{1/3}. \label{34}
\end{equation}

From Eqs. (\ref{26})-(\ref{28}) one has:
\begin{eqnarray}
q_1'&=&1-p_0-\frac{5}{2}\left(\kappa-\frac{2}{3}\right)p_0^2-\frac{25}{8}
\left(\kappa^2-\frac{14}{5}\kappa+\frac{224}{225}\right)p_0^3,\label{35}\\
\mu_1''&=&\mu_1^0(1+3\kappa^2p_0^2), \label{36}\\
\mu_2''&=&\mu_1^0\left(1-2(\kappa-1)p_0+\frac{1}{3}(6\kappa^2+4\kappa-1)p_0^2
\right),\label{37}
\end{eqnarray}
where
\begin{equation}
p_0=\left(\frac{2n_2}{15\kappa^2}\right)^{1/2}. \label{38}
\end{equation}

Using Eqs. (\ref{31})-(\ref{38}), it may be easily shown that
\begin{equation}
\Delta E=E_a-E_b=\frac{1}{2}\hbar\omega_1 N_1\mu_1^0\frac{1-\kappa}{\kappa}
n_2. \label{39}
\end{equation}

From Eq. (\ref{39}) one can see that for $\kappa > 1,$ $\Delta E < 0$, so
the configuration "a" is stable.

Let us now consider the case $n_1 \ll 1$. Approximate solution for
the configuration "a" is given by:
\begin{eqnarray}
q'&=&1-x-\frac{5(3-2\kappa)}{6\kappa}x^2-\frac{(224\kappa^2-630\kappa
+225)}{72\kappa^2}x^3, \label{40}\\
\mu_1'&=&\mu_2^0\beta^{4/5}\left(1+2\frac{(\kappa-1)}{\kappa}x-
\frac{(\kappa^2
-4\kappa-6)}{3\kappa}x^2\right), \label{41}\\
\mu_2'&=&\mu_2^0\beta^{4/5}\left(1+\frac{3}{\kappa^2}x^2\right). \label{42}
\end{eqnarray}
Here
\begin{equation}
x=\left(\frac{2\kappa^2 n_1}{15}\right)^{1/2}. \label{43}
\end{equation}

In the configuration "b" solution has the form;
\begin{eqnarray}
q_1'&=&y+\frac{(\kappa-2/5)}{3\kappa}y^3-\frac{5}{6}y^4, \label{44}\\
\mu_1''&=&\beta^{4/5}\mu_2^0\left[\kappa-(\kappa-1)y^2+\kappa y^3-
\frac{2(\kappa-1)(5\kappa-2)}{15\kappa}y^4\right], \label{45}\\
\mu_2''&=&\beta^{4/5}\mu_2^0(1+y^3), \label{46}
\end{eqnarray}
where
\begin{equation}
y=\left(\frac{2\kappa n_1}{5}\right)^{1/3}. \label{47}
\end{equation}

The energy difference is:
\begin{equation}
\Delta E=\frac{1}{2}\hbar \omega_1 N_2 \mu_2^0 n_1(1-\kappa). \label{48}
\end{equation}

From Eqs. (\ref{39}) and (\ref{48}) it is seen that the configuration "a"
has lower energy if $\kappa=\sqrt{(a_{11}m_2)/(a_{22}m_1)}>1$. For $m_1=m_2$
this is consistent with the qualitative assertion and experimental
observation that it is
energetically favorable for the atoms with the larger scattering length
to form a lower-density shell about the atoms with the smaller
scattering length \cite{[10],pu}.

To evaluate $\Delta E$ in general case it is worth first to estimate
the energy of the phase boundary which arises due to the gradient terms
omitted in the TFA. The surface energy per unit area, the surface tension,
is defined as $\sigma=E_s/S$, where $E_s$ is the surface energy, and $S$
is the interface area. $\sigma$ may be written in the form \cite{[13],[14]}:
\begin{equation}
\sigma=\frac{\hbar\omega_1}{2\sqrt{2}a_{\perp}^2}\left(
\frac{a_{12}}{\sqrt{a_{11}a_{22}}}-1\right)^{1/2}(u_1u_2N_1N_2)^{1/4}
|\psi_1'||\psi_2'|(N_1|\psi_1'|^2+N_2|\psi_2'|^2)^{1/2}. \label{49}
\end{equation}
Taking into account that the surface  area of the ellipsoid with
the semiaxis $a_{\perp}q$ has the form:
\begin{equation}
S=2\pi a_{\perp}^2 q^2\left(1+\frac{1}{\lambda\sqrt{\lambda^2-1}}
\log\frac{1}{\lambda-\sqrt{\lambda^2-1}}\right), \label{50}
\end{equation}
one can estimate the contribution of the surface energy $E_s=\sigma S$
to the total energy of each configuration. To be specific, we will use
the parameters corresponding to the  experiments on $^{87}Rb$ atoms.
In this case $m_1=m_2$, $a_{\perp}=2.4\times 10^{-4} cm$,
$N=N_1+N_2=0.5\times10^6$ atoms.

In Fig. 1(a) we show
the energies of configurations "a" and "b"
(including the surface energy) $E_a/(\hbar \omega_1N)$ (solid line)
and  $E_b/(\hbar \omega_1N)$ (dashed line) as functions of $\log_{10}(n_2)$.
One can see that $E_a$ is always lower than $E_b$. Fig. 1 (b) represents
the difference $\Delta E=(E_a-E_b)/(\hbar \omega_1N)$. The behavior of
$\Delta E$ for small and large values of $n_2$ is well described by
Eqs. (\ref{39}) and (\ref{48}). Fig. 1 (c) illustrates the behavior of
the surface energy as a function of $n_2$. It should be noted that
the surface energy is much smaller
than the interaction energy because the scattering lengths $a_{ij}$
have very close values (see Eq. (\ref{49})).

Let us now consider a trap rotating with frequency $\Omega$ along the
$z$-axis.

For vortex excitation with angular momentum $\hbar l_j$ , the condensate
wave function is given by
\begin{equation}
\psi_{l_{j}}({\bf r})=|\psi_{l_{j}}({\bf r})|e^{il_j\phi}. \label{58}
\end{equation}

In a rotating frame the energy functional of the system is
\begin{equation}
E_{rot}(l_1,l_2)=E(\psi_{l_{1}},\psi_{l_{2}})+\int\,d^3 r
(\psi_{l_{1}}^*+\psi_{l_{2}}^*)i\hbar\Omega \partial_\phi
(\psi_{l_{1}}+\psi_{l_{2}}). \label{59}
\end{equation}

After substituting the wave function for the vortex excitation (\ref{58}) in
Eq. (\ref{59}),
the effective confinement potential for the bosons becomes
$l_1^2\hbar^2/2m_1\rho^2+
l_2^2\hbar^2/2m_2\rho^2+V_1+V_2$, where $V_i=m_i\omega_i(\rho^2+
\lambda^2 z^2)/2$ and $\rho^2=x^2+y^2$. So within the TFA the density
of the vortex state, in separate regions that they do not overlap,
has the form:
\begin{eqnarray}
|\psi_1'({\bf r}')|^2&=&\frac{1}{u_1}\left(\mu_1'(l_1)-(\rho'^2+\lambda^2z'^2)
-\frac{l_1^2}{\rho'^2}\right)
; \label{60} \\
|\psi_2'({\bf r}')|^2&=&\frac{1}{u_2\beta^2}
\left(\mu_2'(l_2)-(\rho'^2+\lambda^2z'^2)-\frac{\beta^2l_2^2}{\rho'^2}\right)
. \label{61}
\end{eqnarray}
The important new qualitative feature of a vortex in the TFA is the
appearance of a small hole of radius $\xi_i$, $\xi_i^2\propto
l_i^2/\mu_i(l_i)$, but the remainder of the condensate density
is essentially unchanged. The fractional change in the chemical
potentials caused by the vortex $(\mu_i'(l_i)-\mu_i')/\mu_i'$
can be shown to be small \cite{[3],[6]}, of the order of
$1/N^{4/5}$. In the calculation of physical quantities involving
the condensate density it is sufficient to retain the no-vortex
density and simply cut off any divergent radial integrals at the
appropriate core sizes $\xi_1^2=l_1^2/\mu_1'$ or
$\xi_2^2=\beta^2l_2^2/\mu_2'$. Note that using the unperturbed density for
calculation of the vortex properties corresponds to the hydrodynamic limit.

In the case of the phase segregated condensate, one finds
from Eqs. (\ref{58}-\ref{59}) and (\ref{15}-{16}) that the energy
change due to the presence of the vortices
$\Delta E=E_{rot}(l_1,l_2)-E_{rot}(0,0)$ has the form:
\begin{eqnarray}
\Delta E&=&\Delta E_{N_1}+\Delta E_{N_2}=\nonumber\\
&=&\frac{1}{2}\hbar \omega_1 N_1 \int\,d_3 r'
\left(\frac{l_1^2}{\rho'^2}|\psi_1'|^2-
\frac{2\Omega l_1}{\omega_1}|\psi_1'|^2
\right)+\nonumber\\
&+&
\frac{1}{2}\hbar \omega_1 N_2 \int\,d_3 r'
\left(\frac{l_2^2\beta^2}{\rho'^2}|\psi_2'|^2-\frac{2\Omega
l_2}{\omega_1} |\psi_2'|^2 \right). \label{62} \end{eqnarray} In
the hydrodynamic limit $\psi_i'$ is given by Eqs. (\ref{12}) and
(\ref{13}).

Let us consider the stable configuration "a".
In the hydrodynamic limit the
location of the phase boundary is given by Eq. (\ref{19}). From (\ref{62})
one has:
\begin{eqnarray}
\frac{\Delta E_{N_1}}{\frac{1}{2}\hbar \omega_1N_1}&=&\frac{5l_1^2
(\mu_1')^{3/2}}{(\mu_1^0)^{5/2}}\left\{\left[\ln\frac{2\mu_1'}{l_1}-\frac{4}{3}
\right]-
\frac{3}{2}q'\left[\left(1-\frac{1}{3}
q'^2\right)\ln\frac{2\mu_1'q'}{l_1}-
\left(1-\frac{q'^2}{9}\right)\right]\right\}-
\frac{2\Omega l_1}{\omega_1}. \label{63}\\
\frac{\Delta E_{N_2}}{\frac{1}{2}\hbar \omega_1N_2}&=&\frac{15l_2^2
(\mu_1')^{1/2}q'}{2(\mu_2^0)^{5/2}}\left[\left(\mu_2'-\frac{1}{3}\mu_1'q'^2
\right)\ln\frac{2\sqrt{\mu_1'\mu_2'}q'}{l_2\beta}-\left(\mu_2'-
\frac{\mu_1'q'^2}{9}\right)\right]-\frac{2\Omega l_2}{\omega_1}. \label{64}
\end{eqnarray}

The critical angular velocities required to produce the vortex states in each
condensate can be determined from the conditions $\Delta E_{N_1}<0,
\Delta E_{N_2}<0$ and have the form:
\begin{eqnarray}
\frac{\Omega_{N_1}}{\omega_1}&=&\frac{5l_1(\mu_1')^{3/2}}{2(\mu_1^0)^{5/2}}
\left\{\left(\ln\frac{2\mu_1'}{l_1}-\frac{4}{3}\right)-\frac{3}{2}q'
\left[\left(1-\frac{q'^2}{3}\right)\ln\frac{2\mu_1'q'}{l_1}-\left(1-
\frac{q'^2}{9}\right)\right]\right\}, \label{65}\\
\frac{\Omega_{N_2}}{\omega_1}&=&
\frac{15l_2(\mu_1')^{1/2}q'}{4(\mu_2^0)^{5/2}}
\left[\left(\mu_2'-\frac{1}{3}\mu_1'q'^2\right)\ln
\frac{2\sqrt{\mu_1'\mu_2'}q'}{l_2\beta}-
\left(\mu_2'-\frac{\mu_1'q'^2}{9}\right) \right]. \label{66}
\end{eqnarray}

Let us consider the asymptotic behavior of the critical angular velocities
(\ref{65}) and (\ref{66}) for $N_2\ll N_1$ and $N_1 \ll N_2$. Using the
approximate solutions (\ref{31})-(\ref{34}), we obtain:
\begin{eqnarray}
\frac{\Omega_{N_1}}{\omega_1}&=&
\frac{5l_1}{2\mu_1^0}\left(\ln
\frac{2\mu_1^0}{l_1}-\frac{4}{3}\right)-
\frac{15l_1}{4\mu_1^0}\left(\ln
\frac{2\mu_1^0q_0}{l_1}-1\right)q_0, \label{67}\\
\frac{\Omega_{N_2}}{\omega_1}&=&
\frac{3l_2\beta^2}{2\mu_1^0q_0^2}\left[
\ln\left( \frac{2\mu_1^0q_0}{\kappa^{1/2}l_2\beta}\right)-1\right] -
\frac{l_2\beta^2}{\mu_1^0}\left[
\ln\left( \frac{2\mu_1^0q_0}{\kappa^{1/2}l_2\beta}\right)\left(1-\frac{4}{5}
\kappa\right)+\frac{7}{12}\kappa-\frac{3}{4}\right]. \label{68}
\end{eqnarray}
From (\ref{67}) and (\ref{68}) one can see that
if $n_2\rightarrow 0$ the critical angular
velocity of the external condensate $\Omega_{N_1}$ tends to that of the
pure condensate
with the scattering length $a_{11}$ (see Eq. (26) in Ref. \onlinecite{[6]}).
The critical angular velocity of the inner
condensate $\Omega_{N_2}$ tends to infinity as
$n_2\rightarrow 0$. However, this consideration can not be applied
to rapidly
rotating gases with $\Omega$ comparable to $\omega_1$ where the form of
the condensate depends on $\Omega$ \cite{[15]}.

In the opposite limit $n_1=N_1/N_2\ll 1$ the critical angular velocities
can be written as
\begin{eqnarray}
\frac{\Omega_{N_1}}{\omega_1}&=&
\frac{l_1}{2\mu_2^0\beta^{4/5}}\ln\frac{2\mu_2^0\beta^{4/5}}{l_1}+
\frac{l_1}{3\kappa\mu_2^0\beta^{4/5}}\left[(3-\kappa)
\ln\frac{2\mu_2^0\beta^{4/5}}{l_1}+3-2\kappa\right]x, \label{69}\\
\frac{\Omega_{N_2}}{\omega_1}&=&
\frac{5l_2\beta^{6/5}}{2\mu_2^0}\left(\ln\frac{2\mu_2^0}{\l_2\beta^{1/5}}
-\frac{4}{3}\right)+
\frac{15l_2\beta^{6/5}}{2\mu_2^0\kappa^2}
\left(\ln\frac{2\mu_2^0}{\l_2\beta^{1/5}}-1\right)x^2. \label{70}
\end{eqnarray}
In deriving (\ref{69}-{\ref{70}) we used the approximate
solutions (\ref{40}-\ref{43}). Note that when $n_1\rightarrow
0$, the critical angular velocity (\ref{70}) has the same form as
the critical velocity for the pure condensate with scattering
length $a_{22}$, $\mu_0\beta^{4/5}$ being the chemical
potential.

Fig. 2 shows the critical angular velocities for the external
($\Omega_{N_1}$) and the inner ($\Omega_{N_2}$) condensates as functions of
$n_2=N_2/N_1$ for $l_1=l_2=1$.

Using Eqs. (\ref{59}) and (\ref{62}), one can find the vortex configurations
which correspond to the energy minimum for a given angular velocity
$\Omega/\omega_1$. In Table 1 we represent angular momenta of condensates
which correspond to the minimum of the total energy
$E_{tot}/\hbar\omega_1$ of
the system for different values of the angular velocity and $n_2$.
In calculation we use
the parameters for the $^{87}Rb$. $E_{tot}$ is calculated as a sum of
$E$ (Eq. (\ref{14}), the surface energy $E_s$ (Eqs. (\ref{49}-{50})),
$\Delta E_{N_1}$ (Eq. (\ref{63})) and $\Delta E_{N_2}$ (Eq. (\ref{64})).

To summarize, we have shown that
in the case of $^{87}Rb$ condensate the atoms in the
state 1 form a shell about the atoms in the state 2,
the critical angular velocity for each state being
drastically dependent on the mutual concentrations.
The critical angular velocity of the outer
component is less than the angular velocity of the inner one.
When the ratio of the number of state 2
atoms to the number of state 1 atoms is small enough
the critical angular velocity of the inner state
becomes very large: in the framework of the hydrodynamic
approximation it turns out to be larger than
the oscillator frequency characterizing the confining
potential and to decrease smoothly with increasing
number of atoms in the state 2. From the Table 1 one can see
that atoms in the state 1 can form a rotating ring around
the resting core of the atoms in the state
2.

This work was supported in part by NATO Grant No.PST.CLG.976038.
V.N.R and E.E.T acknowledge the financial support from the
Russian Science Foundation through the Grant No.98-02-16805.
STC was partly supported by NASA under grant No. CRG8-1427.

\begin{table}
\caption{
\label{table1}}
\begin{tabular}{lcccr}
$\Omega/\omega_1$&$n_2$&$l_1$&$l_2$&$E_{tot}$\\
\tableline
$0.1$&$1.0$&$1$&$0$&$1.296863\times10^7$\\
$0.1$&$10.0$&$2$&$0$&$1.287497\times10^7$\\
$0.15$&$1.0$&$2$&$0$&$1.295151\times10^7$\\
$0.2$&$1.0$&$3$&$1$&$1.292344\times10^7$\\
\end{tabular}
\end{table}

\begin{figure}
\caption{
(a) The total energies of configurations "a" and "b"
$E_a/(\hbar \omega_1N)$ (solid line)
and  $E_b/(\hbar \omega_1N)$ (dashed line)
as functions of $\log_{10}(n_2)$.
(b) The difference $\Delta E=(E_a-E_b)/(\hbar \omega_1N)$
as a function of $\log_{10}(n_2)$.
(c) The surface energies as functions of $\log_{10}(n_2)$.
Solid line corresponds to the surface energy $E_{sa}$
for the configuration "a",
dashed line - to the surface energy $E_{sb}$ for the configuration "b".
}
\end{figure}
\begin{figure}
\caption{
Critical angular velocities $\Omega_{N_1}/\omega_1$ and
$\Omega_{N_2}/\omega_2$ as functions of $\log_{10}(n_2)$ for
the configuration "a". Solid line corresponds to the outer condensate 1,
dashed line - to the inner condensate 2.
}
\end{figure}
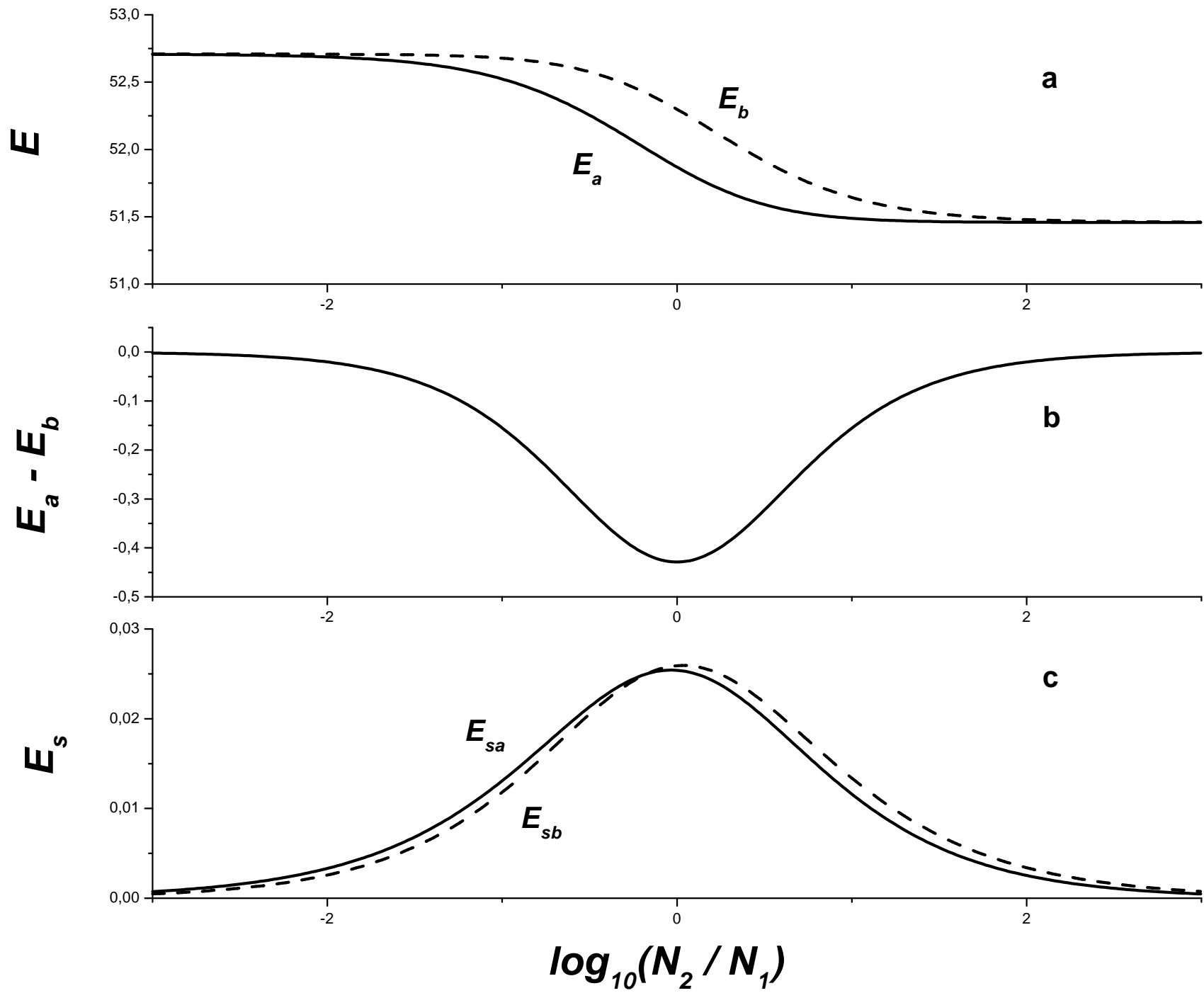
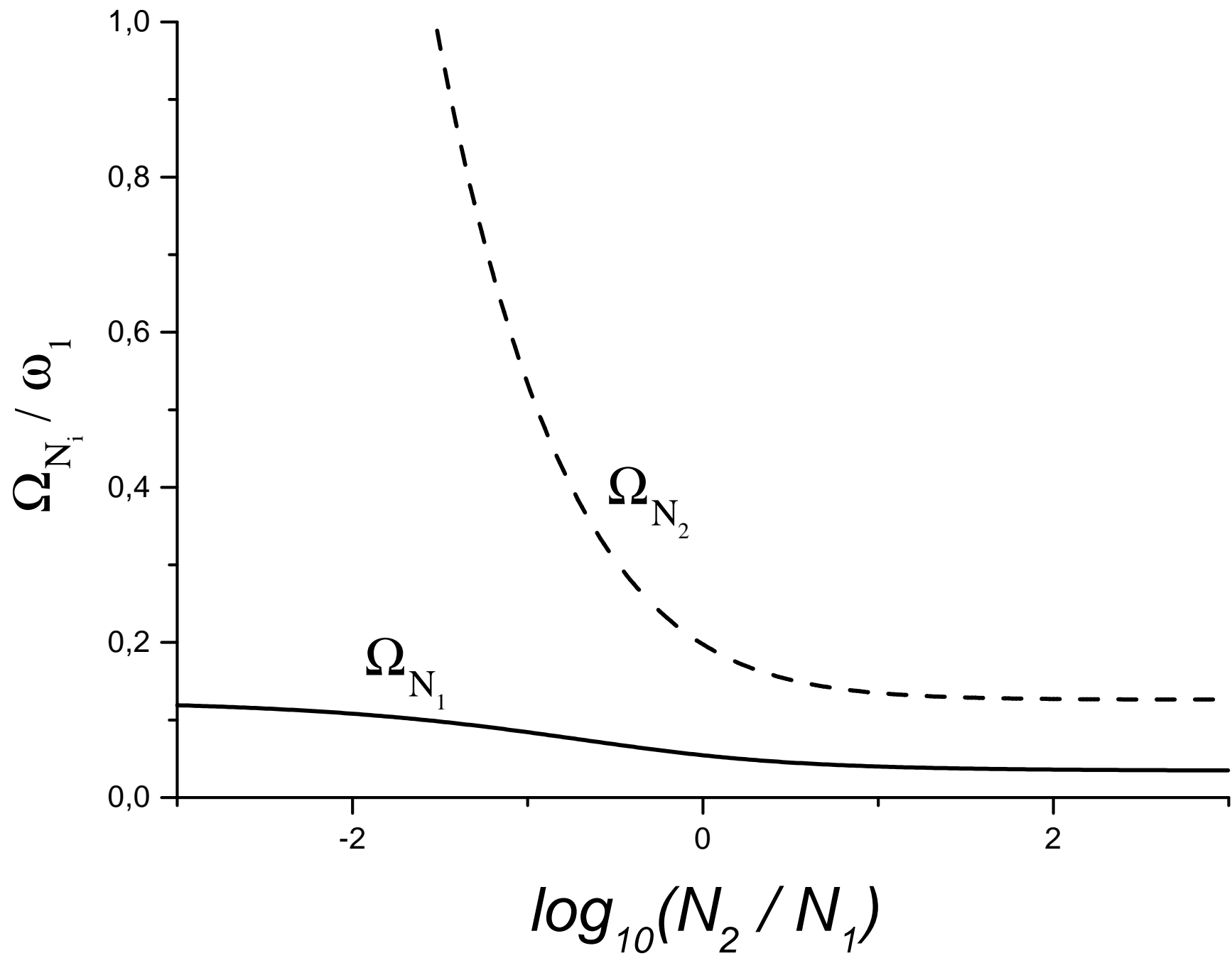
\end{document}